# Optical "Shorting Wires"

*Andrea Alù and Nader Engheta\**

*University of Pennsylvania, Department of Electrical and Systems Engineering,*

*Philadelphia, PA, U.S.A.*

**Abstract**

Connecting lumped circuit elements in a conventional circuit is usually accomplished by conducting wires that act as conduits for the conduction currents with negligible potential drops. More challenging, however, is to extend these concepts to optical nanocircuit elements. Here, following our recent development of optical lumped circuit elements, we show how a special class of nanowaveguides formed by a thin core with relatively large (positive or negative) permittivity surrounded by a thin concentric shell with low permittivity may provide the required analogy to "wires" for optical nano-circuits.

The fundamental concepts for extending the classic circuit theory to nanocircuits at optical frequencies have been introduced and developed by our group in a recent letter [1]. In that work, we have shown that it may be possible to realize lumped nanoinductors, nanocapacitors and nanoresistors by utilizing sub-wavelength nanoparticles with negative-real (plasmonic), positive-real (non-plasmonic) and imaginary (lossy) permittivity, respectively. In this framework, the role of conduction current density flowing in classic circuits is replaced by the displacement current density $J_d = -i\omega\varepsilon E$ circulating in such a nanocircuit, with $\omega$ being the operating radian frequency under an $e^{-i\omega t}$ time convention and $\varepsilon$ the local permittivity. While for a single element it has been proven that the classic Kirchhoff laws in the circuit theory are satisfied in this nanocircuit paradigm [1] in terms of the current $I$ defined as the flux integral of $J_d$ across the optical nanoelement and the optical potential $V$ between its two "ends", it is more challenging to envision a complex nanocircuit containing several closely packed nanocircuit elements connected with a special pattern. Even though we have shown that the basic parallel and series combinations of two optical nanoelements are indeed possible by suitably positioning two interconnected particles with respect to the orientation of the external field [2], and that the optical nanotransmission lines may be designed following simple circuit considerations on periodic arrangements of such elements [3]-[4], the flexibility and possibility of connecting relatively distant optical lumped nanoelements or in connecting insulated nanoparticles that are closely spaced on a nanocircuit board is not as straightforward as in a classic circuit at lower frequencies [5]. This is because at lower frequencies, at which classic circuits rely on the conduction current ($J_c = \sigma E$, with $\sigma$ being the local conductivity and $E$ the local electric field) circulating across the elements, the background material (i.e., air) or the substrate of a circuit board are naturally very poorly conducting ($\sigma \simeq 0$), and therefore the elements are connected with each other only through the shorting metallic conducting wires attached to their terminals. At high frequencies (infrared and visible), however, the displacement current $J_d$ on which our nanocircuits rely may also flow in the surrounding background materials, in some sense connecting all nanocircuit nanoelements together in an unwanted manner. In this nanocircuit paradigm, the role of material conductivity is indeed replaced by the local material permittivity $\varepsilon$. It follows, as discussed in [5], that the equivalent of a poorly conducting material in the framework of our optical nanocircuit theory is represented by a low-permittivity ($\varepsilon$-near-zero, ENZ) material, for which $J_d$ is close to zero despite the non-zero values of the local electric field. An ENZ material surrounding a lumped nanocircuit element would indeed "insulate" the element from the unwanted coupling with other neighboring elements, at least for what concerns the displacement current flow in it [5]. Analogously, the role of nanoconnectors may be taken, in this nano-circuit analogy, by large permittivity materials ($\varepsilon$-very-large, EVL, either positive or negative large permittivity), which may "connect" relatively distant optical nanoelements with a small potential drop, since for a given level of current $J_d$ the corresponding longitudinal electric field, and the consequent voltage drop, are very low across the EVL connector. Following these heuristic concepts [1]-[5], the representation of a complex nanocircuit system is getting closer to realization with feasible materials and technology at optical frequencies, i.e., employing naturally available plasmonic, non-plasmonic and/or polaritonic materials. In order to make the picture more complete, however, it may be desirable to envision an optical "wire" that may connect two parts of an optical nanocircuit or two or more nanocircuit elements, in the same manner as classic metallic conducting wires at lower frequencies allow flow of conduction currents with no current leakage and with negligible voltage drop. Such an optical "wire" should be able to pass the displacement current flow with essentially no phase delay, without spreading and leaking it around its path, and with a sufficiently low optical potential drop across its terminals. In the following, we investigate the possibility of providing these functionalities of an optical wire in the optical nanocircuit domain, exploiting the anomalous properties of EVL and ENZ materials.

We may heuristically envision the optical "wire" as a deeply sub-wavelength cylindrical waveguide that is formed by a core with an EVL material surrounded by a thin ENZ shell, which bounds the displacement current flow in the EVL core and does not allow current leakage across it. Here, we verify this prediction by studying this problem analytically and numerically, demonstrating how a displacement current flow with a low optical potential drop and low phase delay may be indeed obtained with such a structure. Since ENZ and EVL materials exist at infrared/optical frequencies [6] or they may be realized as layered metamaterials [7], realization of such optical nanowaveguides may be envisioned in the optical domains.

Consider the geometry of Fig. 1a: a cylindrical core of permittivity $\varepsilon_{EVL}$ surrounded by a shell with permittivity $\varepsilon_{ENZ}$ and surrounded by a background material with permittivity $\varepsilon_0$. The corresponding radii are, respectively, $a_{core}$ and $a_{shell}$. All the material permeabilities are assumed to be the same as that of free space $\mu_0$.

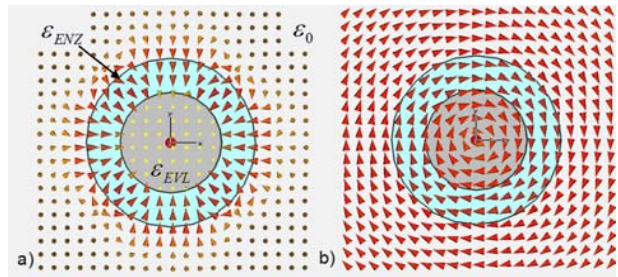

Figure 1 – (Color online) The cross-section of an optical wire composed of an EVL core and an ENZ shell. Electric (a) and magnetic (b) field distribution, as calculated from full-wave simulations [11] and consistent with Eq. (1) for a nanowire with $a_{core} = 25\,nm$, $a_{shell} = a_{core}/0.6$, $\varepsilon_{ENZ} \simeq (0.01 + i\,0.01)\varepsilon_0$, $\varepsilon_{EVL} \simeq 50\varepsilon_0$ at a background wavelength of $\lambda_0 = 500\,nm$. Arrows are drawn in scale and darker colors correspond to larger field amplitudes.

An azimuthally symmetric TM mode supported by such a waveguide has the following field distributions in each region of space in Fig. 1 [8]:

$$\mathbf{H} = \hat{\boldsymbol{\varphi}}\, A\, z_1\left(\sqrt{k^2 - \beta^2}\,\rho\right) e^{i\beta z}$$
$$\mathbf{E} = \hat{\boldsymbol{\rho}}\, A\, \frac{\beta}{\omega\varepsilon}\, z_1\left(\sqrt{k^2 - \beta^2}\,\rho\right) e^{i\beta z} + \qquad , \qquad (1)$$
$$+ i\,\hat{\mathbf{z}}\, A\, \frac{\sqrt{k^2 - \beta^2}}{\omega\varepsilon}\, z_0\left(\sqrt{k^2 - \beta^2}\,\rho\right) e^{i\beta z}$$

in the corresponding cylindrical reference system of Fig. 1, where $\beta$ denotes the propagation constant of a guided mode, $z_n$ is the suitable $n$-th order Bessel function ($J_n$ in the core region, a linear combination of $J_n$ and $Y_n$ in the shell region and the Hankel function $H_n^{(1)}$ in the background region). Moreover, $\varepsilon$ is the local permittivity in the different regions, $k = \omega\sqrt{\varepsilon\mu_0}$ is the local wave number and $A$ a suitable coefficient (different in each region) to match the continuity of the tangential components of fields at the two interfaces. Fig. 1 shows an example of these field distributions.

The general dispersion relation for the modes guided by this structure with arbitrary values for material permittivities is well known [8]. Only for $\beta > k_0 = \omega\sqrt{\varepsilon_0\mu_0}$ this dispersion equation admits real solutions, which are the surface waves supported by a core-shell cylinder. In the limit of $|\varepsilon_{EVL}| \gg \varepsilon_0$ and $|\varepsilon_{ENZ}| \ll \varepsilon_0$, together with the deep sub-wavelength cross-sectional dimensions $a_{shell} \ll \lambda_0 = 2\pi/k_0$, the eigen-solutions assume the following interesting closed-form expression:

$$\beta^2 = \frac{k_{ENZ}^2}{2\ln a_{shell}/a_{core}}\left[i\pi - 2\frac{J_0(k_{EVL}a_{core})}{k_{EVL}a_{core}J_1(k_{EVL}a_{core})} - \ln\left(\frac{k_0^2 a_{core}^2}{4}\right) - 2\gamma\right], \tag{2}$$

where $\gamma$ is the Euler constant.

This closed-form expression is quite remarkable in its utility to find the propagation constants of guided modes supported by the waveguide of Fig. 1 in the limits of interest here, and it provides many physical insights into its anomalous guidance properties. First of all, it is evident how the values of $\beta$ are proportional to the wave number in the shell region $k_{ENZ}$, which tends to zero for low-epsilon shell materials. This is physically expected, due to the low variation of phase in the ENZ material, and is consistent with the findings in [9], regardless of the field excited inside the EVL core. Due to its low value, $\beta$ is required to be necessarily complex in order to take into account the radiation leakage. In other words, since $\beta < k_0$, the supported modes are leaky-modes, and thus they necessarily radiate in the background region as manifested in the intrinsic complex nature of $\beta$ given in Eq. (2). (Note the analogies with the structure described in [10]).

The similarities with a standard shorting metallic conducting wire in a radio-frequency circuit become evident for the traveling modes described by (2): the phase variation along the wire is almost zero (since $\beta \propto k_{ENZ} \simeq 0$) and the electric and magnetic field distributions around it, as depicted in Fig. 1, closely resemble those of a regular conducting wire at radio frequencies. It should be noted that a regular metallic wire also radiates some small energy when a time-varying current flows along its path, and its field distribution is distributed in the background material as described in (1). Usually the radiation losses are negligibly small in classic RF circuits, due to the very small electrical size of RF circuits and the subsequent quasi-static nature of the problem. Eq. (2) can fully take into account the radiation associated with a displacement current propagating with $\beta \propto k_{ENZ} \simeq 0$ along this optical nanowire, again consistent with its lower frequency counterpart.

If we now consider the limit $\varepsilon_{ENZ} = 0$, we will note that the ENZ shell will act as a perfect magnetic conductor for this TM polarization, reducing the field to zero outside the core and imposing the stringent dispersion relation $J_1\left(\sqrt{k_{EVL}^2 - \beta^2} a_{core}\right) = 0$ on the EVL core radius and/or material. This condition, however, requires a very high positive value of $\varepsilon_{EVL}$, due to the sub-wavelength dimensions of $a_{core}$. An anomalous leaky-mode with $\beta = 0$ may be envisioned in this limit, whose properties however go beyond the interest of the present letter.

Here we consider the scenario that assumes reasonable realistic values for ENZ materials, for which $\varepsilon_{ENZ}$, while still small, is necessarily distinct from zero due to frequency dispersion and the inherent presence of losses. In this case Eq. (2) holds, which may provide more degrees of freedom for the spectrum of possible guided modes supported by this waveguide. Inspecting Eq. (2), we note that for a propagating mode to be supported ($\text{Re}[\beta] \neq 0$) we just require our geometry to satisfy the following condition:

$$2\frac{J_0(k_{EVL}a_{core})}{k_{EVL}a_{core}J_1(k_{EVL}a_{core})}+\ln\left(\frac{k_0^2 a_{core}^2}{4}\right)+2\gamma<0, \tag{3}$$

which imposes some constraints on $\varepsilon_{EVL}$ for a fixed dimension of the inner core. (We note how, in this limit, the outer radius of the shell $a_{shell}$ is not crucial in ensuring the guidance properties of Eq. (2), for which condition (3) is sufficient, even though it may play a role in the amplitude of $\beta$, appearing in the denominator of Eq. (2)). The imaginary part of $\beta$, in this limit and under condition (3), is intrinsically small, since it is proportional to $k_{ENZ}$. Its value, and the corresponding radiation from the wire, becomes negligible when the left-hand of Eq. (3) becomes sufficiently negative to overcome the term $(i\pi)$ in (2), which happens near the condition $J_1(k_{EVL}a_{core})=0$, i.e., when the ENZ shell acts similar to a magnetic boundary.

Applying (3), it is found that for small values of $a_{core}/\lambda_0$ there is a relatively large range of possible values of $\varepsilon_{EVL}$ to have a guided mode in the structure, reasonably smaller than those required in principle if the ENZ shell were replaced by an ideal magnetic boundary. For instance, for an inner core with $a_{core}=\lambda_0/20$, feasible at optical frequencies, a guided mode is supported in the region $13.17\varepsilon_0<\varepsilon_{EVL}<148.8\varepsilon_0$. Figure 2, as an example, shows the calculated modal dispersion for a waveguide with $a_{core}=\lambda_0/20$, $a_{shell}=a_{core}/0.8$, varying the material permittivities. Several interesting features are evident in these plots: first of all, consistent with the previous discussion and fairly independent of the value of $\varepsilon_{ENZ}$, the region of guidance is delimited by Eq. (3), which ensures a low imaginary part of $\beta$ and a fast-wave propagation (low phase variation along the nano-wire) over a broad range of values for $\varepsilon_{EVL}$. It is interesting to note, however, as evident from Fig. 2, that this equation may be satisfied also by negative values of $\varepsilon_{EVL}$, allowing to employ plasmonic materials in the core region, like noble metals and polar dielectrics or semiconductors, widely available in nature at infrared and optical frequencies. The asymptotes in the figure coincide with the roots of the dispersion relation $J_1(k_{EVL}a_{core})=0$, that a small waveguide of the same dimensions with magnetic boundaries would support. As previously noticed, near these asymptotes at the end of each branch of guided modes $\text{Im}[\beta]$ and the corresponding radiation losses tend to zero.

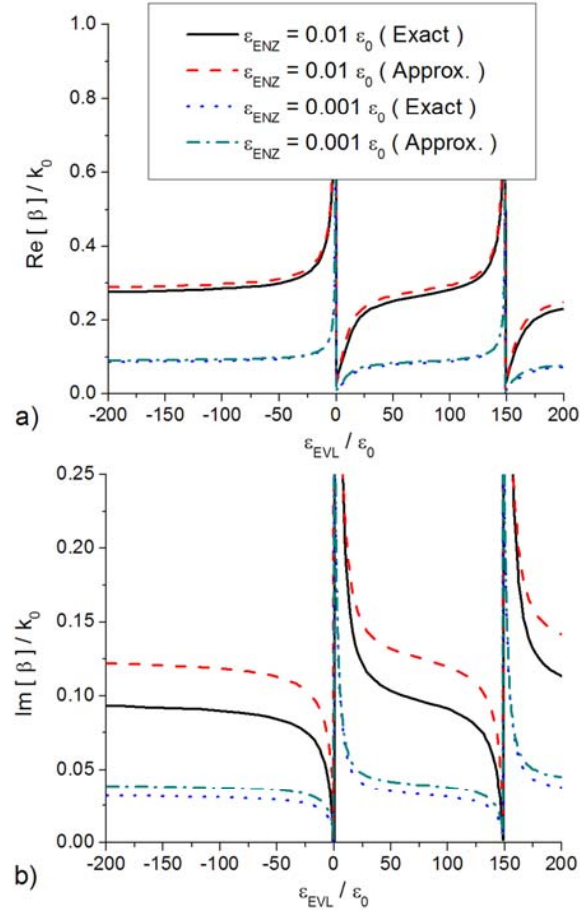

Figure 2 – (Color online) Dispersion of the guided wave number as a function of $\varepsilon_{EVL}$ varying $\varepsilon_{ENZ}$ for

$$a_{shell} = a_{core} / 0.8.$$

The dependence of the wave number over the geometrical parameters of the waveguide is well described by the approximate expression (2), as the comparison between the exact and approximate lines show in Fig. 2. The variation of the dispersion curves for a thinner or thicker shell, i.e., varying $a_{shell}$, is again well described by Eq. (2), which implies a simple scaling of the corresponding value of $\beta$ as the $\ln a_{shell} / a_{core}$. This implies that a thicker ENZ core reduces both the phase variation along the wire and the radiation losses, as expectable from the previous physical considerations, due to the better insulation provided by the shell. The same result is obtained by reducing $\left|\varepsilon_{ENZ}\right|$, as seen from the figure, and consistent with physical considerations. The nano-wire operation is indeed appealing, both for positive and negative values of $\varepsilon_{EVL}$, ensuring a relatively broad range of values of permittivities for which the required anomalous guidance may be obtained. Inspecting the results of Fig. 2b concerning the imaginary part of the guided wave number $\beta$, it is evident how the damping of such modes due to radiation losses is relatively low, ensuring that in the sub-wavelength scale of an optical nano-circuit these nano-wires may transport the displacement current from one end to the other, with small energy loss (in Fig. 2 we have not considered the presence of material losses, which may enter into play when the nano-wire has a very sub-wavelength cross-section or when the imaginary part of permittivity is relatively high). The results of Fig. 2b, however, are substantially unchanged when reasonable ohmic losses for optical materials are considered, as we show in the following examples that consider realistic material losses. As an aside, the sign of $\text{Im}[\beta]$ in Fig. 2 also ensures that the guided modes in that specific example are all forward, independent of the sign of $\varepsilon_{EVL}$. It may be proven, more in general, that the sign of $\varepsilon_{ENZ}$ is the one to affect directly the direction of propagation of this anomalous mode: backward propagation is obtained when $\varepsilon_{ENZ}$ has a negative real part (i.e., before the plasma frequency of the material).

This may be seen more in detail in Fig. 3, where the full dispersion of the guided modes for a nano-wire with $a_{shell} = a_{core}/0.8$ has been plotted considering the necessary dispersion of the involved plasmonic materials. In this example, the ENZ shell has indeed been assumed to be plasmonic in nature with a Drude-like dispersion $\varepsilon_{ENZ}(\omega) = \left[1 - \omega_p^2/(\omega(\omega + i\Gamma))\right]\varepsilon_0$, with $\omega_p = 2\pi \cdot 600\,THz$ and $\Gamma = 10^{-3}\omega_p$. In this way, we model reasonable dispersion and losses for a plasmonic material at optical frequencies [6]. In the figure three different possibilities are considered for the EVL core material: a positive $\varepsilon_{EVL}$, and two different negative values for the core permittivity. In these latter cases the materials have also been assumed to have a Drude frequency dispersion with corresponding ohmic losses. The values indicated in the figure insets correspond to the permittivity values at $f_p = 600\,THz$. In the three cases we obtain nano-wire behavior over a relatively wide range of frequency centered at the plasma frequency of the shell material. The calculated low guided wave number $\text{Re}[\beta]$ and the low damping factor $\text{Im}[\beta]$ ensure the feasibility of the nano-wire for optical nano-circuit applications. It is noted that a positive $\varepsilon_{EVL}$ guarantees a wider region of low $\text{Re}[\beta]$, but slightly higher damping factors away from the plasma frequency of the ENZ material. A more negative permittivity for the core material, still in the range of realistic values of permittivity at optical frequencies, as in the example with $\text{Re}[\varepsilon_{EVL}] = -100$, show better performance, as expected. However, reasonable performance over a relatively wide bandwidth is also obtained with a less negative permittivity. It should be noted how, for always positive $\text{Im}[\beta]$, the sign of $\text{Re}[\beta]$ flips together with the sign of $\text{Re}[\varepsilon_{ENZ}]$ at the plasma frequency $f_p$. This is consistent with the previous discussion on the possibility of forward and backward modes in this regime.

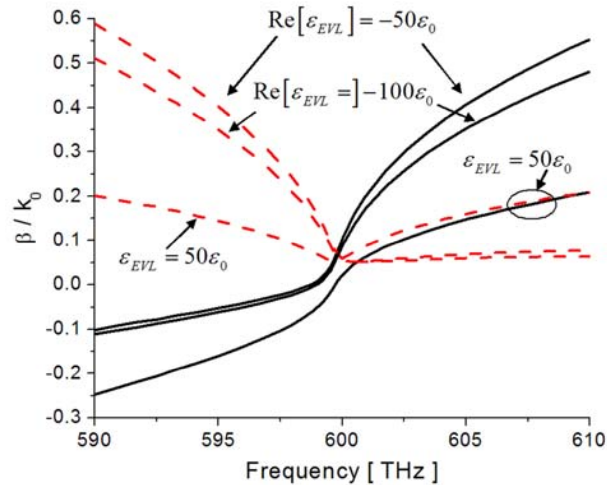

Figure 3 – (Color online) Dispersion of the guided wave number for a nanowire with $a_{shell} = a_{core}/0.8$ considering the frequency dispersion of plasmonic materials. Solid black lines refer to the real part of $\beta$ and dashed red lines to its imaginary part.

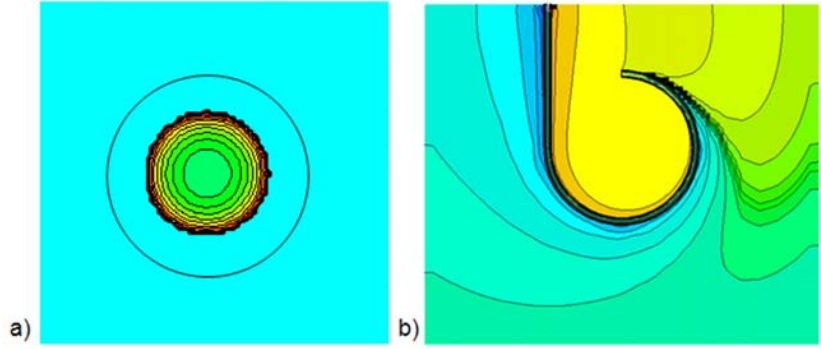

Figure 4 – (Color online) Full-wave simulations [11] of: (a) distribution of the optical displacement current density on a transverse cross section of the nanowire; (b) snapshot in time of the optical magnetic field distribution for a bent nanowire with the geometry of Fig. 1 (the length of the straight section is $\lambda_0$ and the radius of curvature of the bend is $\lambda_0/2$). Darker (brighter) colors correspond to higher values.

Figure 4a shows the displacement current density $-i\omega\varepsilon_{EVL}E_{core}$ across a cross section of the nano-wire of Fig. 1 obtained using numerical simulations, and Fig. 4b presents the magnetic field distribution (a snapshot in time domain) for a bent geometry of such a wire. Since in the RF regime, a metallic wire can still operate as intended, even when it is bent, here we show that this optical "wire" still functions with its interesting properties, when it is bent. Both figures reveal interesting features of such an optical "wire", consistent with the findings described above. In Fig. 4a it is evident how the displacement current is all concentrated in the inner core of the nano-wire, similar to a classic shorting conducting metallic wire at lower frequencies. This is also confirmed by the field distributions reported in Fig. 1, which shows a small longitudinal electric field component in the core, and which resemble those of a classic wire carrying a uniform current. The ENZ shell in this geometry indeed allows confining the displacement current inside the core with a very low optical potential drop (We note that despite the displacement current being high in the core region, the corresponding longitudinal electric field, which contributes to the potential drop along the wire, is very small due to the high value of permittivity of the core material).

Fig. 4b shows the full-wave simulation of a bent nanowire, demonstrating the striking guidance properties of such an optical "wire". It can be seen how the mode is tightly guided around the sub-wavelength waveguide with a very low phase variation along the line. Despite the electrical length of the bent nano-wire shown in Fig. 4b, which is much longer than what would be required in nano-circuit applications, the phase delay along the wire is still negligible and the phase follows the nano-wire despite its bend. Also the calculated potential drop along the nanowire is substantially lower than the one expected along a similar electrical length in free-space or along a regular waveguide, since the distribution of longitudinal electric field in the inner region is substantially lower than the transverse electric field in the surrounding free-space (consistent with Fig. 1a).

To conclude, we have reported how the anomalous guiding properties of an EVL thin cylindrical nanowire surrounded by a thin ENZ plasmonic shell may act as an optical "wire" with interesting potentials in the framework of optical nanocircuit theory and nanoelectronics. The analysis presented here has also derived a closed-form solution for the complex guided wave numbers of this optical wire. Various methods for realization of this optical wire may be forecasted: one way may consist of the use of sub-wavelength stacks of plasmonic and non-plasmonic materials, which, depending on the orientation of the field, may act as effective ENZ or EVL materials (see, e.g., [7]). Another possibility may be by utilizing a relatively high permittivity substrate: since the present analysis has been developed relative to a generic background permittivity $\varepsilon_0$, it is expected that the use of a relatively high-permittivity background may reasonably relax the requirements of a low permittivity for the ENZ shell.

This work is supported in part by the U.S. Office of Naval Research (ONR) grant number N 00014 - 07-1-0622